%% file: Bdecay_EPJA_v2.tex
\documentclass[epj]{svjour}
\usepackage{amssymb,amsmath,epsfig,bm,color,slashed,mathrsfs}
\usepackage{amsmath,bm}
\usepackage{txfonts}
\usepackage[square,numbers]{natbib}
\usepackage{graphicx}
\usepackage{multirow}
\usepackage[colorlinks=true,linkcolor=blue,citecolor=cyan,urlcolor=blue]{hyperref}
\usepackage{nicefrac}
\usepackage{exscale} 
\usepackage{gensymb}
\usepackage{slashed}
\usepackage{color}
\newcommand{\bce}{\begin{center}} 
\newcommand{\be}{\begin{equation}}
\newcommand{\bea}{\begin{eqnarray}}
\newcommand{\eea}{\end{eqnarray}}

\newcommand{\vecj}{{\bm j}}

\newcommand{\ie}{\textit{i.e.,~}}
\newcommand{\eg}{\textit{e.g.,~}}
\newcommand{\vecE}{{\bm E}}
\newcommand{\vecB}{{\bm B}}
\newcommand{\vecnabla}{{\bm \nabla}}

\def\apj{ApJ}%
\def\aap{A\&A}%
\def\mnras{MNRAS}%
\def\prd{Phys.~Rev.~D}%
\definecolor{red}{rgb}{0.8,0,0}
\definecolor{violet}{rgb}{0.4,0,0.4}
\definecolor{green}{rgb}{0,0.5,0.0}
\definecolor{navy}{rgb}{0.0,0.0,0.6}
\definecolor{orange}{rgb}{0.8,0.2,0.0}
\definecolor{blue}{rgb}{0.3,0.0,0.8}

\begin{document}
\title{Electrical resistivity and Hall effect in binary neutron-star mergers} 

\author{ 
          Arus~Harutyunyan\inst{1} \and 
          Antonios~Nathanail\inst{1} \and
          Luciano~Rezzolla\inst{1,2} \and 
          Armen Sedrakian\inst{2,1} 
}

\institute{ Institute for Theoretical Physics,
  J. W. Goethe-University, Max-von-Laue Str.\ 1, D-60438 Frankfurt am
  Main, Germany \and Frankfurt Institute for Advanced Studies,
  Ruth-Moufang str.\ 1, D-60438 Frankfurt am Main, Germany }

\abstract{ 
We examine the range of rest-mass densities, temperatures and magnetic
fields involved in simulations of binary neutron-star mergers and
identify the conditions under which the ideal-magnetohydrodynamics
approximation breaks down and hence the magnetic-field decay should be
accounted for. We use recent calculations of the conductivities of
warm correlated plasma in envelopes of compact stars and find that the
magnetic-field decay timescales are much larger than the
characteristic timescales of the merger process for lengthscales down
to a meter. Because these are smaller than the currently available
resolution in numerical simulations, the ideal-magnetohydrodynamics
approximation is effectively valid for all realistic simulations. At
the same time, we find that the Hall effect can be important at low
densities and low temperatures, where it can induce a non-dissipative
rearrangement of the magnetic field. Finally, we mark the region in
temperature and density where the hydrodynamic description breaks
down.
} 
\date{\today}
\maketitle

\section{ Introduction}  

The recent detections of gravitational waves by the LIGO and Virgo
detectors have opened a new chapter in multimessenger astronomy. Among
these observations, the recent detection of a gravitational wave
signal GW170817 originating from a binary neutron star (NS)
inspiral~\cite{Abbott2017} and a gamma-ray burst by the Fermi
satellite GRB170817A~\cite{Abbott2017d}, which were followed by
electromagnetic counterparts~\cite{Abbott2017b, Abbott2017c,
  Coulter2017, Smartt2017}, has demonstrated the prominent role that
binary neutron-star inspirals can play in astrophysics, particle
physics, and cosmology~\cite{Abbott2017h}. 

The observations of GW170817 and GRB170817A provided the first direct
evidence that a class of short gamma-ray bursts can be associated with
the inspiral and merger of binary compact stars. As the magnetic field
plays a central role in the generation of the GRBs, the physics of
inspiral and merger of magnetized neutron stars can now be constrained
directly by observations. The magnetohydrodynamics (MHD) simulations
of these processes have advanced steadily over the recent past and the
post-merger massive compact object can now be followed up to the point
of collapse to a black hole over timescales of the order of tens of
milliseconds; for reviews see, \eg \cite{Faber2012:lrr,
  Paschalidis2016, Baiotti2016}. General-relativistic simulations have been carried
out in the ideal MHD limit (\ie with infinite-conductivity) by several
groups \cite{Rezzolla:2011, Kiuchi2014, Palenzuela2015, Kawamura2016,
  Ruiz2016,Kiuchi2017,Ruiz2018} and a number of works have included the resistive (\ie
finite conductivity) effects~\cite{Dionysopoulou:2012pp,Palenzuela2013b,Palenzuela2013a,Dionysopoulou2015}.

The aim of this work is to investigate the conditions under which the
non-ideal MHD effects can be important in the contexts of neutron star
mergers and the evolution of the post-merger object. The motivation
for doing so is three-fold. Firstly, the past resistive MHD
simulations \cite{Dionysopoulou:2012pp, Dionysopoulou2015} have used
simplified functional forms of conductivities aimed to provide a
smooth interpolation between highly conducting core and low-conducting
magnetosphere; these were independent of the temperature and
composition of matter and were functions of its rest-mass density
only. Finite-temperature, composition-dependent conductivities for
relevant low-densities have become available
recently~\cite{Harutyunyan:2016a} and will be used in our estimates
below.  Secondly, the relevance of the Hall effect which arises in
anisotropically conducting plasma has not been assessed in the context
of binary neutron star merger physics; for studies in the context of
cold neutron stars see~\cite{Gourgouliatos2016,Kitchatinov2017}.  We
will show below that the Hall effect can play an important role in a
certain density-temperature regime. Thirdly, the density-temperature
regime where the MHD approximation breaks down and a kinetic
description of plasma is needed has not been established so far.

The novel features of the post-merger remnant object have recently
motivated a study of related transport properties of dense matter,
such as thermal conduction and bulk and shear viscosities in the
absence of electromagnetic fields~\cite{Alford2018}. On the basis of
simple estimates supported by numerical simulations of non-dissipative
merger and post-merger process, it was concluded that bulk viscous
effects could be important. Along similar lines, here we will adopt a
semi-analytical approach to estimate the various timescales with
processes listed above, \ie Ohmic dissipation, Hall effect and
breakdown of MHD. This will be supported by the results of {\it ideal}
MHD simulations, which allow us to estimate the relevant scales of
gradients of magnetic field components and density. These simulations
validate the fact that the magnetic field gradients are much smaller
than those for the density - a crucial underlying assumption of our
semi-analytical analysis.  Thus, our work is aimed at aiding future
dissipative MHD simulations in identifying the various physical
regimes when realistic conductivities are employed.

This paper is organized as follows. In Sec.~\ref{sec:BNS_simulations}
we discuss the ideal MHD simulations of binary neutron star mergers.
Section~\ref{sec:Ohmic_Hall} collects the relevant formulae for 
the timescales involved. Our numerical results for the time- and
lengthscales are presented in Sec.~\ref{sec:numerics}. Finally, our
conclusions are  given in Sec.~\ref{sec:conclusions}.

\begin{figure*}[t]
  \centering 
  \includegraphics[width=0.95\textwidth]{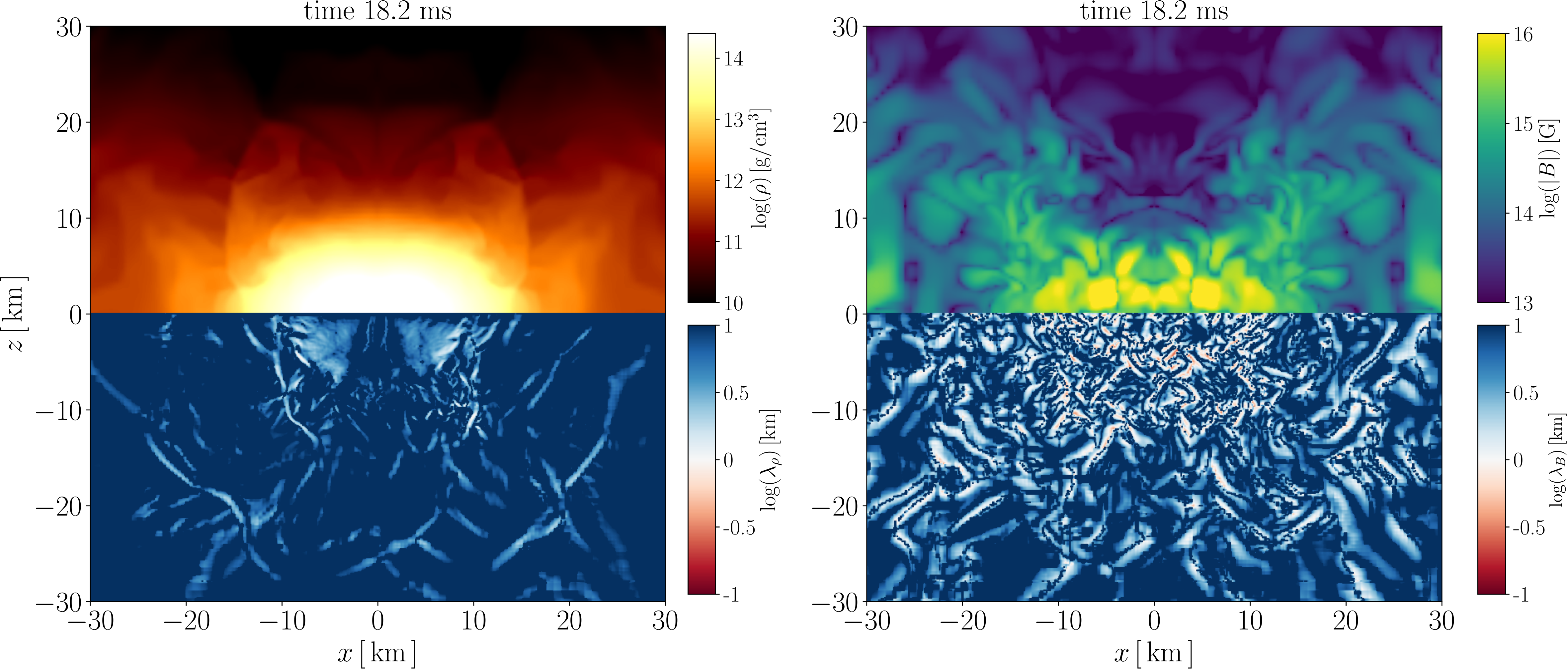}
  \caption{Two-dimensional cuts in the $(x, z)$ plane of the rest-mass
    density $\rho$ (\textit{left panel}) and of the modulus of the
    magnetic field $|B| = (B_iB^i)^{1/2}$ (\textit{right panel}); the
    snapshots refer to $t-t_{\rm mg} \simeq 14.7\, {\rm ms}$ of a
    representative magnetized binary. Shown in the lower parts of the two
    panels are the scale-heights of the two quantities, $\lambda_{\rho}
    \approx \sqrt{(\rho/\partial_x \rho)^2 + (\rho/\partial_z\rho)^2}$
    and $\lambda_{B} \approx \sqrt{(B/\partial_x B)^2 +
      (B/\partial_zB)^2}$; note that $\lambda_{\rho}=\mathcal{O}(10\,{\rm
      km})$ and $\lambda_{B}=\mathcal{O}(1\,{\rm km})$, on average.}
  \label{fig:BNS}
\end{figure*}

\section{Binary neutron-star merger simulations} 
\label{sec:BNS_simulations}

The aim of this section is to discuss the relevant scales of magnetic
field components and thermodynamical parameters in the simulations of
magnetized binary neutron star mergers. As shown below the onset of
the resistive MHD regime depends on the {\it resolution} of
simulations, therefore we will, in addition, discuss this scale. In
view of continuous progress in resolving smaller scales in the
simulations, it is important to identify the scales at which the
resistive MHD becomes mandatory irrespective of current computing
capabilities. The current state of the art with respect to resolution
is as follows.  

The magnetic field in a merger process can be substantially amplified
via the Kelvin-Helmholtz instability (KHI)~\cite{Rasio99}.  The
magnetorotational instability (MRI) \cite{Velikhov1959,
  Chandrasekhar1960} can also take place in the post-merger
object. Both instabilities are normally not captured at the typical
resolutions of global simulations, that are limited to a grid
resolution of the order of $100-200\,{\rm m}$ \cite{Anderson2008,
  Liu:2008xy, Giacomazzo:2009mp, Kiuchi2009, Giacomazzo:2010}. At the
same time, simulations at ultra-high resolutions of $70 \,{\rm m}$
\cite{Kiuchi2015a} or even $12 \,{\rm m}$ \cite{Kiuchi2017} struggle
to describe the KHI in a convergent regime and hence determine
reliably the actual magnetic-field amplification. Similarly, while
less severe requirements are needed to capture the MRI
\cite{Siegel2013, Kiuchi2017}, the role played by parasitic
instabilities in limiting the amplification of the magnetic field is
still unclear \cite{Rembiasz2016}. Finally, the local shearing-box
simulations of special-relativistic MHD turbulence seem to indicate
that equipartition level magnetic field magnitudes could be reached
\cite{Obergaulinger10, Zrake2013}.

Let us now turn to the lengthscale over which significant variations
of the magnetic field take place, which we denote as
$\lambda_B$. Clearly, this lengthscale is bound from below by the
resolution scale of any given simulation, $l_{0}$, which was already
mentioned just above. We note that the decay timescale of the magnetic
field is proportional to the square of the lengthscale of variation of
the magnetic field (see below), therefore it is limited from below
by the spatial resolution used in the numerical
simulation. Characteristic timescales of simulations also set the
upper bound $\tau_0$ over which the relevant quantities need to change
to be relevant dynamically. The current highest resolution
simulations of the merger process, which were aimed at a better
resolution of the KHI and MRI and monitoring of magnetic
field amplification \cite{Siegel2013, Kiuchi2015, Kiuchi2017}, are
limited in time after merger due to numerical costs. Lower-resolution
simulations can be carried out up to the point of the collapse to a
black hole or the formation of a differentially rotating and unstable
neutron star, and last typically tens of milliseconds.

To extract the relevant lengthscales we used a simulation of the merger
of an equal-mass magnetized binary neutron-star system with a total ADM
mass of $M_{\rm ADM}= 3.25 M_{\odot}$ and initial orbital separation of
$45\,{\rm km}$. For the sake of reproducibility, we use the open-source
Illinois GRMHD code \cite{Etienne2015} with the finest resolution of 
$l_0 \approx 227\,{\rm m}$. Each star has a baryon mass of $1.625
M_{\odot}$ and is endowed initially with a poloidal magnetic field fully
contained inside the star. The maximum magnetic field in the initial
configuration is a bit less than $10^{16}\, {\rm G}$ and the merger takes
place at $t_{\rm mg} \simeq 3.5\, {\rm ms}$ after the start of the
simulation.

For illustrative purposes, we show in Fig.~\ref{fig:BNS} the rest-mass
density (left panel) and the magnetic-field configuration (right panel)
of the hypermassive neutron star (HMNS) created by the merger, at time
$t-t_{\rm mg} \sim 14.7\,{\rm ms}$. Similar behaviours are seen
also at earlier and later times. Note that the rest-mass density of
matter drops from nuclear saturation density down to $\rho\simeq
10^{10}\, {\rm g\, cm}^{-3}$ over the lengthscale of $30\,{\rm km}$, with
sharp gradients at $z\simeq 10$ and $x\simeq 20\, {\rm km}$. Except for
this transition region, the rest-mass density profile is smooth over the
lengthscales of the order of several kilometres. In other words, an
approximately constant rest-mass density can be considered when making
estimates of quantities over lengthscales of the order of $10\,{\rm
  km}$. This is shown in the lower part of the left panel of
Fig.~\ref{fig:BNS}, which reports in a colormap the average scale-height
of the rest-mass density $\lambda_{\rho}:= \rho/(\nabla \rho) \approx
\sqrt{(\rho/\partial_x \rho)^2 + (\rho/\partial_z\rho)^2}$; clearly, the
large dominance of the dark-blue color in the low-density regions outside
the core of the HMNS, where the low conductivity of matter may induce
resistive-MHD effects, indicates that $\lambda_{\rho} \gtrsim
\mathcal{O}(10\,{\rm km})$ there.

The right panel in Fig.~\ref{fig:BNS} shows the magnitude of the magnetic
field, which shows a filament structure with small-scale variations
having characteristic lengthscale of the order of
kilometres. Interestingly, even in the lowest-density regions, where the
rest-mass density is quite uniform, there are significant structures in
the field, indicating substantial variations over the lengthscale of
kilometres. This is shown in the lower part of the right panel of
Fig.~\ref{fig:BNS}, which reports the average scale-height of the
magnetic field $\lambda_{B}:=B/(\nabla B) \approx \sqrt{(B/\partial_x B)^2 +
  (B/\partial_zB)^2}$; clearly, in this case, $\lambda_{B} \gtrsim
\mathcal{O}(1\,{\rm km})$ in the low-density regions.

In summary, ideal-MHD simulations generically indicate that:
\textit{(a)} the characteristic lengthscales over which magnetic-field
variations can be significant are of the order of $1\,{\rm km}$ or
less, with the lower limit on this scale obviously given by the
resolution of the simulation; \textit{(b)} the characteristic
lengthscales over which the rest-mass density variations in the
low-density regions can be significant are of the order of
$10\,{\rm km}$. The density of matter is approximately constant over
the lengthscales of variation of the magnetic field; \textit{(c)} the
characteristic timescales relevant for the simulations are of the
order of $10\,{\rm ms}$. These considerations now motivate our
assessment of conditions of applicability of ideal MHD that we will
provide next and substantiate some of the approximations we will
adopt.

\section{Ohmic and Hall timescales} 
\label{sec:Ohmic_Hall}

The electrical conductivity of dense matter in neutron stars was
studied extensively in the cold regime for temperatures
$T\le 0.1\,{\rm MeV}$ which are relevant for isolated neutron stars
and neutron stars in X-ray
binaries~\cite{2015PhRvL.114c1102H,2008ApJ...677..495I,1999A&A...351..787P,
  1998PhRvL..81.5556B,1993ApJ...418..405I,
  1993ApJ...404..268I,1984ApJ...285..758I,
  1984MNRAS.209..511N,1984ApJ...277..375M, 1983ApJ...273..774I}; for a
review see \cite{Schmitt2017}. In this regime, the matter is strongly
degenerate and the ionic component is solidified, \ie the electrons
are scattered by phonons and impurities.  The physical properties of
the product of the merger, which may form a long-lived but unstable
configuration, differ markedly from ordinary neutron stars; for
example, the temperatures are of the order of some tens of MeV
($\sim 10^{11}\,{\rm K}$) \cite{Palenzuela2015, Hanauske2016,
  Kastaun2017, Bovard2017}. Furthermore, their differential rotation
may lead to additional dissipative heating of the post-merger object
before it loses its stability and collapses into a black
hole~\cite{Kastaun2014, Hanauske2016}.

The conductivity tensor in a magnetic field for applications to
neutron star mergers was obtained recently~\cite{Harutyunyan:2016a,
  Harutyunyan:2016b}.  This study provides the conductivity tensor for
temperatures extending up to $T \simeq 10\,{\rm MeV}$, rest-mass
densities down to $\rho \simeq 10^6\, {\rm g\, cm}^{-3}$ and for
non-quantizing magnetic field $B\le 10^{14}\,{\rm G}$ . In
particular, the anisotropy in the conduction due to the field, \ie the
tensor structure of the conductivity was taken fully into account.

Now we turn to the derivation of the time- and lengthscales of
interest in terms of the conductivity of matter.  As is well-known, the MHD
description of low-frequency phenomena in neutron stars is based on
the following Maxwell equations
\begin{eqnarray}
\label{eq:Maxwell_eqs}
   \vecnabla \times \vecE=-\frac{1}{c}\frac{\partial\vecB}{\partial t},\qquad
   \vecnabla \times \vecB=\frac{4\pi}{c}\vecj\,,
\end{eqnarray}
where we assume that the magnetic permeability of matter is unity and
neglect the displacement current. The system of equations
\eqref{eq:Maxwell_eqs} should be closed by Ohm's law for the electric
current $\vecj$. In the presence of a strong magnetic field, the
conduction in neutron star crusts is anisotropic and Ohm's law should be
written in a tensor form, \ie $\vecj=\bm{\hat{\sigma}}\vecE$,
where $\bm{\hat{\sigma}}$ is the electrical conductivity tensor. Substituting
the current in the second equation of~\eqref{eq:Maxwell_eqs} according to Ohm's 
law and eliminating the electric field $\vecE$, we obtain the induction 
equation, which describes the magnetic-field evolution as
\begin{eqnarray}
  \label{eq:mag_field}
  \frac{\partial {\bm B}}{\partial t}=
  -{\vecnabla \times} ~\left(\bm{\hat{\varrho}}  \, \, {\vecnabla
    \times}{\bm B}\right)\,,
\end{eqnarray}
where $\bm{\hat{\varrho}}$ is the electrical resistivity tensor (or
magnetic viscosity tensor) and is simply the inverse of the electrical
conductivity tensor, \ie $ \bm{\hat{\varrho}} =
({c^2}/{4\pi}){\bm{\hat{\sigma}}}^{-1}$.

\subsection{Isotropic conductivity tensor}
\label{sec:isotropic_case}

In the case of isotropic conduction $\hat\sigma_{ij} =
\delta_{ij}\sigma={\rm const.}$, and Eq.~\eqref{eq:mag_field} reduces
to
\begin{eqnarray}
  \label{eq:mag_field_is}
  \frac{4\pi}{c^2}\frac{\partial \vecB}{\partial t}=
  -{\vecnabla \times} \left(\frac{{\vecnabla \times}\vecB}{\sigma}\right)\,.
\end{eqnarray}
If, as found in Sec.~\ref{sec:BNS_simulations}, the characteristic
lengthscales of variation of the magnetic field are smaller than the
lengthscales over which the rest-mass density varies and since
$\sigma$ is a function of the rest-mass density, we can approximate
$\sigma$ as constant over the scale height of the magnetic field. We
stress that the validation of this approximation is based on fully
dynamical ideal-MHD simulations of binary neutron star mergers
described in Sec.~\ref{sec:BNS_simulations}.  Then, upon taking into
account that
${\vecnabla \times}\, ({\vecnabla \times}\,\vecB) =\vecnabla\,
(\vecnabla\cdot\vecB)-\Delta \vecB$
and $\vecnabla\cdot\vecB=0$, we obtain from
Eq.~\eqref{eq:mag_field_is}
\begin{eqnarray}
  \label{eq:mag_field_is1}
  \frac{4\pi\sigma}{c^2}\frac{\partial \vecB}{\partial t}=
  \Delta\vecB\,.
\end{eqnarray}
A qualitative estimate of the magnetic field decay timescale $\tau_d$
now can be obtained from Eq.~\eqref{eq:mag_field_is1} if we
approximate $|\Delta \vecB|\simeq B/\lambda_B^2$ and
$|\partial \vecB/\partial t|\simeq B/\tau_d$.  From these estimates,
we find that the magnetic field decay (or diffusion) timescale due to
Ohmic dissipation is given by the well-known expression
\begin{eqnarray}
  \label{eq:decay_time}
\tau_d = \frac{4\pi\sigma \lambda^2_B}{c^2}\, .
\end{eqnarray}
%

\subsection{Anisotropic conductivity tensor}
\label{sec:anisotropic_case}

The electrical conductivity tensor in the presence of a strong magnetic
field can be decomposed as
\begin{eqnarray}
  \label{eq:sigma_kj}
  \sigma_{kj}=\delta_{kj}\sigma_0-\epsilon_{kjm}b_m
  \sigma_1 +b_kb_j\sigma_2\,,
\end{eqnarray}
where $b_k=B_k/B$. Inverting Eq.~\eqref{eq:sigma_kj} we find for
$\bm{\hat{\varrho}}$ 
\begin{eqnarray}
  \label{eq:varrho_kj}
  \varrho_{ik}=\delta_{ik}\varrho_0+\epsilon_{ikm}b_m
  \varrho_1 +b_ib_k\varrho_2\,,
\end{eqnarray}
where its components are given by 
\begin{eqnarray}
  \label{eq:varrho_0}
  \varrho_0&=&\frac{c^2}{4\pi}\frac{\sigma_0}{\sigma_0^2+\sigma_1^2},\qquad
  \varrho_1=\frac{c^2}{4\pi}\frac{\sigma_1}{\sigma_0^2+\sigma_1^2},\\
  \label{eq:varrho_12}
  \varrho_2&=&\frac{c^2}{4\pi\sigma}\frac{\sigma_1^2-\sigma_0\sigma_2}{\sigma_0^2+\sigma_1^2}\,,
\end{eqnarray}
where $\sigma=\sigma_0+\sigma_2$ is the longitudinal 
conductivity,
\ie the electrical conductivity in the absence of magnetic field. If the
magnetic field is directed along the $z$ axis, then the tensors
$\bm{\hat{\sigma}}$ and $\bm{\hat{\varrho}}$ have the form
\begin{eqnarray}
  \label{eq:sigma_varrho_matrix}
\bm{\hat{\sigma}} &=&
\begin{pmatrix}
    \sigma_0 & -\sigma_1 & 0 \\
    \sigma_1 & \sigma_0 & 0 \\
    0 & 0 & \sigma
\end{pmatrix}\,,\\
\bm{\hat{\varrho}} &=& \frac{c^2}{4\pi}\frac{1}{\sigma_0^2+\sigma_1^2}
\begin{pmatrix}
    \sigma_0 & \sigma_1 & 0 \\
    -\sigma_1 & \sigma_0 & 0 \\
    0 & 0 & {(\sigma_0^2+\sigma_1^2)}/{\sigma}
\end{pmatrix}\,.
\end{eqnarray}
In fact, one can write down Drude-type formulas for each of the three
components of the conductivity tensor in the cases of degenerate or
nondegenerate electrons~\cite{Harutyunyan:2016a}
\begin{eqnarray}
  \label{eq:sigma_drude}
  &&\sigma = \frac{n_ee^2c^2\tau}{\varepsilon},\qquad
  \sigma_0 =
  \frac{\sigma}
       {1+(\omega_{c}\tau)^2}\,,\\
  \label{eq:sigma1_drude}     
       &&
       \sigma_1 =\frac{(\omega_{c}\tau) \sigma}
             {1+(\omega_{c}\tau)^2}=(\omega_c\tau)\sigma_0\, .
\end{eqnarray}
 Here $n_e$ is
the electron number density, $e$ is the elementary charge, $\tau$ is the electron mean collision time [or
microscopic relaxation time, see Eqs. (13) and (30) of
\cite{Harutyunyan:2016a}], $\omega_c := ecB\varepsilon^{-1}$ is
the cyclotron frequency, and $\varepsilon$ is the characteristic
energy-scale of electrons.
In the degenerate-electron
regime $\varepsilon=\varepsilon_F$, where $\varepsilon_F$ is the 
electron Fermi energy; for nondegenerate electrons, instead,
${\varepsilon}= 3T/2 + \sqrt{(3T/2)^2+m^2c^4}$, where $T$ is the
temperature and $m$ is the electron mass~\cite{Harutyunyan:2016a}. From the last expression we recover the
well-known results $\varepsilon =mc^2+3T/2$ and $\varepsilon= 3T$ for the nonrelativistic and ultrarelativistic regimes,
respectively.

Although the expressions \eqref{eq:sigma_drude} and \eqref{eq:sigma1_drude} were derived
only for strongly degenerate or nondegenerate electrons, it has been
argued, on the basis of full numerical study, that these can be
applied also for arbitrary degeneracy if one takes for $\varepsilon$
the characteristic (to the regime) energy scale of
electrons~\cite{Harutyunyan:2016a}.

The components of the resistivity tensor are related to those of the
conductivity tensor by relations \eqref{eq:varrho_0} and \eqref{eq:varrho_12}; using 
Eqs.~\eqref{eq:sigma_drude} and 
\eqref{eq:sigma1_drude} we obtain the following estimates 
\begin{eqnarray}
  \label{eq:varrho_0_Drude}
  &&\varrho_0\simeq \frac{c^2}{4\pi\sigma}\equiv\varrho,\\
  \label{eq:varrho_12_Drude}
  &&\varrho_1 
  \simeq \frac{c^2}{4\pi}\frac{\sigma_1}{\sigma\sigma_0}
  \simeq (\omega_c\tau)\varrho
  \simeq\frac{cB}{4\pi n_ee},\qquad
  \varrho_2\simeq 0\,.\quad
\end{eqnarray}

We now estimate the right-hand side of Eq.~\eqref{eq:mag_field} using the
matrix~\eqref{eq:varrho_kj}. Writing
\bea
  \label{eq:rot_B}
  (\bm{\hat{\varrho}}~ {\vecnabla \times}\vecB)_i &=&
  \varrho_0\epsilon_{ijl} \partial_j B_l-\varrho_1
  (b_l \partial_i B_l-b_j \partial_j B_i)\nonumber\\
&+&\varrho_2b_ib_k\epsilon_{kjl} \partial_j B_l\,,
\eea
and approximating again $\partial_i B\simeq B/\lambda_B$ and using expressions
\eqref{eq:varrho_0_Drude} and \eqref{eq:varrho_12_Drude}, we arrive at
\begin{eqnarray}
  \label{eq:A_estimate}
  |\bm{\hat{\varrho}} \,\, {\vecnabla \times}\vecB|\simeq 
  {\rm max}(1,\omega_c\tau) \varrho \frac{B}{\lambda_B}\,.
\end{eqnarray}
In the case of small magnetic field $\omega_c\tau\ll 1$ and we
recover the isotropic case from Eq.~\eqref{eq:mag_field}. The
evolution of magnetic field is then determined by the Ohmic diffusion
timescale $\tau_d$ given by Eq.~\eqref{eq:decay_time}. Conversely, in
the strongly anisotropic regime, where $\omega_c\tau\gg 1$, the
magnetic field evolution is determined by the characteristic timescale
$\tau_B $ given by 
\begin{eqnarray}
  \label{eq:B_time}
\tau_B   = \frac{\tau_d}{\omega_c\tau}
  =\frac{ 4\pi n_ee \lambda^2_B}{cB}
   =\frac{ 4\pi e\rho \lambda^2_B}{cB}\frac{Z}{Am_n},
\end{eqnarray}
which follows from Eqs.~\eqref{eq:mag_field}, \eqref{eq:decay_time}, 
\eqref{eq:varrho_12_Drude} and \eqref{eq:A_estimate}.
In the last step, we used the condition of the charge neutrality, which implies
$n_e = Z\rho/(Am_n)$, where $Z$ and $A$ are the charge and the mass
number of nuclei, respectively, and $m_n$ is the atomic mass unit. 

Thus, in the strongly anisotropic regime (which is realized for
sufficiently high magnetic field and low rest-mass densities) the
characteristic timescale over which the magnetic field evolves is
reduced by a factor of $\omega_c\tau$ due to the Hall effect. We see
from Eq.~\eqref{eq:B_time}, that the timescale $\tau_B$ decreases with
an increase of the magnetic field and is independent of the electrical
conductivity $\sigma$ (and, consequently, also of the temperature),
which is in contrast to the Ohmic diffusion timescale $\tau_d$. The
physical reason for this difference lies in the fact that the Hall
effect {\it per-se} is not dissipative. Note however that it can act
to facilitate Ohmic dissipation. For instance, the Hall effect may
cause the fragmentation of magnetic-field into smaller structures
through the Hall instability, which can then accelerate the decay of
the field via standard Ohmic dissipation [see
\cite{Gourgouliatos2016,Kitchatinov2017} and references therein.]

\begin{figure*}
\begin{center}
\includegraphics[width=0.95\textwidth]{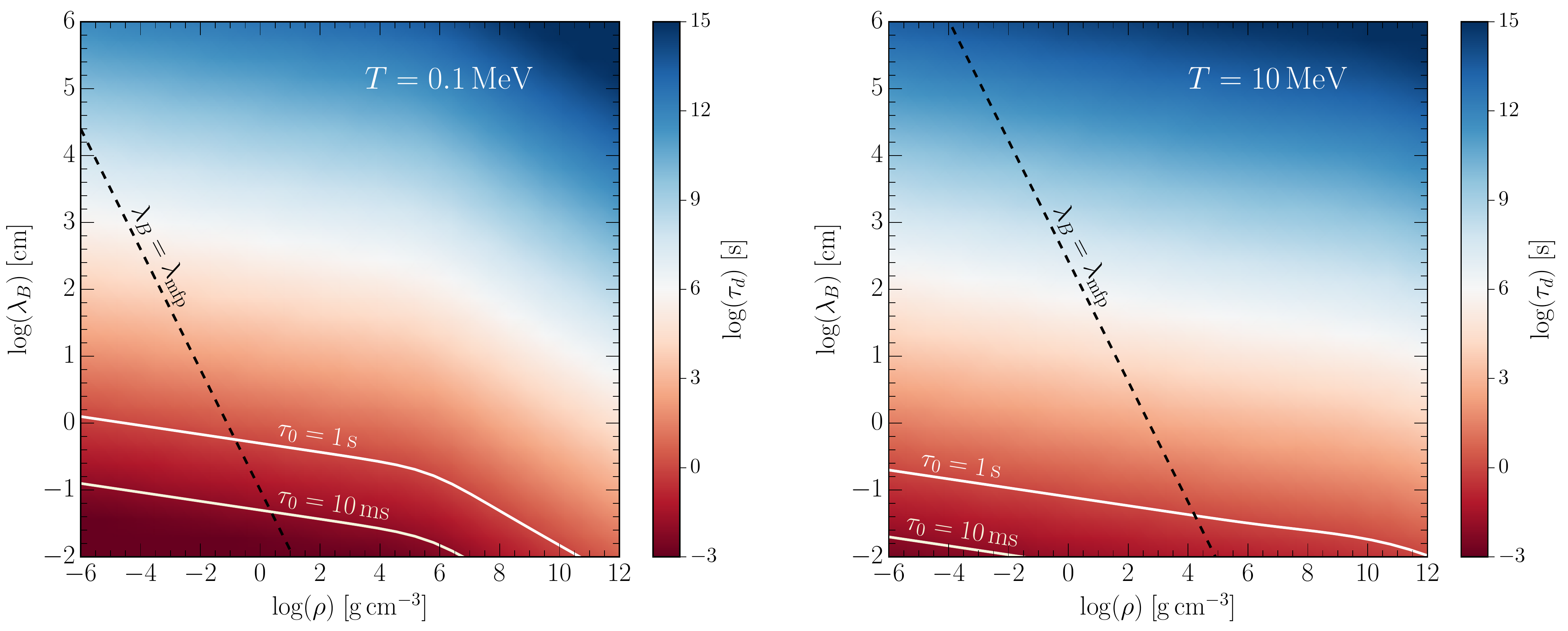}
\caption{Dependence of the magnetic-field decay timescale $\tau_d$ on the
  rest-mass density and the typical scale-height of the magnetic field
  $\lambda_B$ for $T=0.1\,{\rm MeV}\sim 1.2\times 10^{9}\,{\rm K}$
  (\textit{left panel}) and $T=10\,{\rm MeV} \sim 1.2\times10^{11}\,{\rm
    K}$ (\textit{right panel}). The solid lines correspond to typical
  timescales of $\tau_0=10\,{\rm ms}$ and $\tau_0=1\,{\rm s}$,
  respectively. The region with $\lambda_B\geq \lambda_{\rm mfp}$ and
  $\tau_d\leq\tau_0$ is where the Ohmic dissipation becomes
  important. The dashed lines show where $\lambda_B$ becomes equal to the
  electron mean free path $\lambda_{\rm mfp}(\rho,T)$; below these lines
  the MHD description breaks down.}
\label{fig:decay_time_2D}
\end{center}
\end{figure*}

\subsection{Estimating magnetic field decay timescale}

In order to estimate the Ohmic diffusion timescale given by
Eq.~\eqref{eq:decay_time} we use the following fit formula for the
electrical conductivity $\sigma=\sigma(\rho,T,Z)$ ~\cite{Harutyunyan:2016a}
\begin{eqnarray}
  \label{eq:sigma_fit}
  \sigma = \frac{1.5\times 10^{22}}{Z}
  \left(\frac{T_F}{1~{\rm MeV}}\right)^a
  \bigg(\frac{T}{T_F}\bigg)^{-b}
  \bigg(\frac{T}{T_F}+d\bigg)^{c} {\rm s}^{-1}\,,
  \nonumber\\
\end{eqnarray}
where $T$ and $T_F$ are the temperature of the stellar matter and the Fermi
temperature of electrons, respectively. While more accurate tables are
available~\cite{Harutyunyan:2016b}, the expression above is accurate
up to $10\%$, which exceeds the accuracy required for our qualitative
estimates. The rest-mass density dependence of $\sigma$ is via the
Fermi temperature given by
$T_F= 0.511 \big[\sqrt{1+(Z\rho_6/A)^{2/3}}-1\big]\,{\rm MeV}$, 
where $\rho_6 := \rho/(10^6\,{\rm g\  cm}^{-3})$. The
fitting parameters $a,b,c,d$ depend on the charge number of the nucleus via
the formulas~\cite{Harutyunyan2017}
\begin{eqnarray}
  \label{eq:fit_constants1}
  a(Z) &=& 0.924 - 0.009 \log Z + 0.003 \log^2 Z\,,\\
  b(Z) &=& 0.507 - 0.028  \log Z - 0.025 \log^2 Z\,,\\
  c(Z) &=& 1.295 - 0.018   \log Z - 0.022 \log^2 Z\,,\\
  \label{eq:fit_constants2}
  d(Z) &=& 0.279 + 0.056  \log Z + 0.035 \log^2 Z\,.
\end{eqnarray}
Below, we are interested mainly in the low-density regime of stellar
matter, where low values of conductivity could render resistive and
Hall effects important. At such low densities, the matter consists
mainly of hydrogen or helium nuclei. 

Assuming now for simplicity
$Z=A=1$ and using Eqs.~\eqref{eq:sigma_fit}--\eqref{eq:fit_constants2}
we find an approximation for Eq.~\eqref{eq:decay_time}
\begin{eqnarray}
  \label{eq:decay_time1}
  \tau_d(\rho,T)&\simeq &2.1\times 10^{12}\,\,
  \left(\frac{T_F}{1\,{\rm MeV}}\right)\bigg(\frac{T}{T_F}\bigg)^{-0.5}
  \nonumber\\
  &\times &
  \bigg(\frac{T}{T_F}+0.28\bigg)^{1.3}
  \left(\frac{L}{1\,{\rm km}}\right)^2~{\rm s}\,.
\end{eqnarray}
We will use this equation to obtain the low-density estimate
below, see Eq.~\eqref{eq:decay_time2a}.

\section{Numerical results}
\label{sec:numerics}

We start our discussion with the applicability of the ideal-MHD
approximation under conditions relevant for the binary neutron star
merger simulations by comparing $\tau_d$ with $\tau_0$. To set a lower
limit for the former, we need to choose values for the rest-mass
density and the temperature bearing in mind the values encountered in
numerical simulations. More specifically, choosing a rest-mass density
still compatible with the hydrodynamic description 
$\rho_6 \lesssim 1$ (see discussion below),
and a temperature in this regime $T\gtrsim 1\,{\rm MeV}$,
we find $T_F\simeq 0.25\,\rho_6^{2/3}\,{\rm MeV} \ll T$, 
therefore from Eq.~\eqref{eq:decay_time1} we find an estimate
%
\begin{eqnarray}
  \label{eq:decay_time2a}
  \tau_d \simeq 5\times 10^{11}\,\, \left(\frac{\rho}{1~{\rm
      g\,cm^{-3}}}\right)^{0.1} \bigg(\frac{T}{1\,{\rm
      MeV}}\bigg)^{0.8}
  \left(\frac{\lambda_B}{1\,{\rm km}}\right)^2~{\rm s} ,\nonumber\\
\end{eqnarray} 
which is clearly much larger than the typical timescales
$\tau_0 = 10\,{\rm ms}$ involved in a merger. Given the current
limitations on the resolution to the order of a meter and choosing the
most favorable temperature and density values we can obtain an
effective lower limit on $\tau_d $ by substituting in
Eq.~\eqref{eq:decay_time2a} $\lambda_B=1\,{\rm m}$,
$\rho \simeq 10^{-3}\,{\rm g\ cm}^{-3}$, and $T=0.1\,{\rm MeV}$, in
which case $\tau_d\sim 10^{4}\,{\rm s}$. This value effectively sets a
lower limit for $\tau_d$ for the adopted resolution scale and is still
much larger than $\tau_0$\footnote{Smaller rest-mass densities can be
  reached in binary mergers, but for $\lambda_B=1\,{\rm m}$ and
  $T=0.1\,{\rm MeV}$ the MHD approach breaks down already at
  $\rho\lesssim 10^{-3}$~g~cm$^{-3}$; see the discussion in
  Sec.~\ref{sec:MHD_val}.}.  Thus, our analysis suggests that
resistive effects do not play an important role in the MHD
phenomenology of binary neutron star mergers and the ideal-MHD
approximation is applicable in the entire parameter range of
interest. Our conclusion is in contrast with the previous paradigm of
onset of dissipative MHD in low-density regime
\cite{Dionysopoulou:2012pp, Dionysopoulou2015} where the nearly
zero-conductivity of matter would have implied $\tau_d \to 0$.

Figure~\ref{fig:decay_time_2D} displays the timescale $\tau_d$ in a broad
range of the rest-mass density (including the extremely dilute regime
relevant for the stellar atmosphere) and the typical lengthscale of
magnetic field structures for two values of the temperature. It is
seen that at the currently available resolution scale the ideal MHD is
applicable in most of the density-temperature range. The
dissipative effect would become important only if the resolution is
increased by at least several orders of magnitude. 

\begin{figure} 
\begin{center}
\includegraphics[width=0.95\columnwidth]{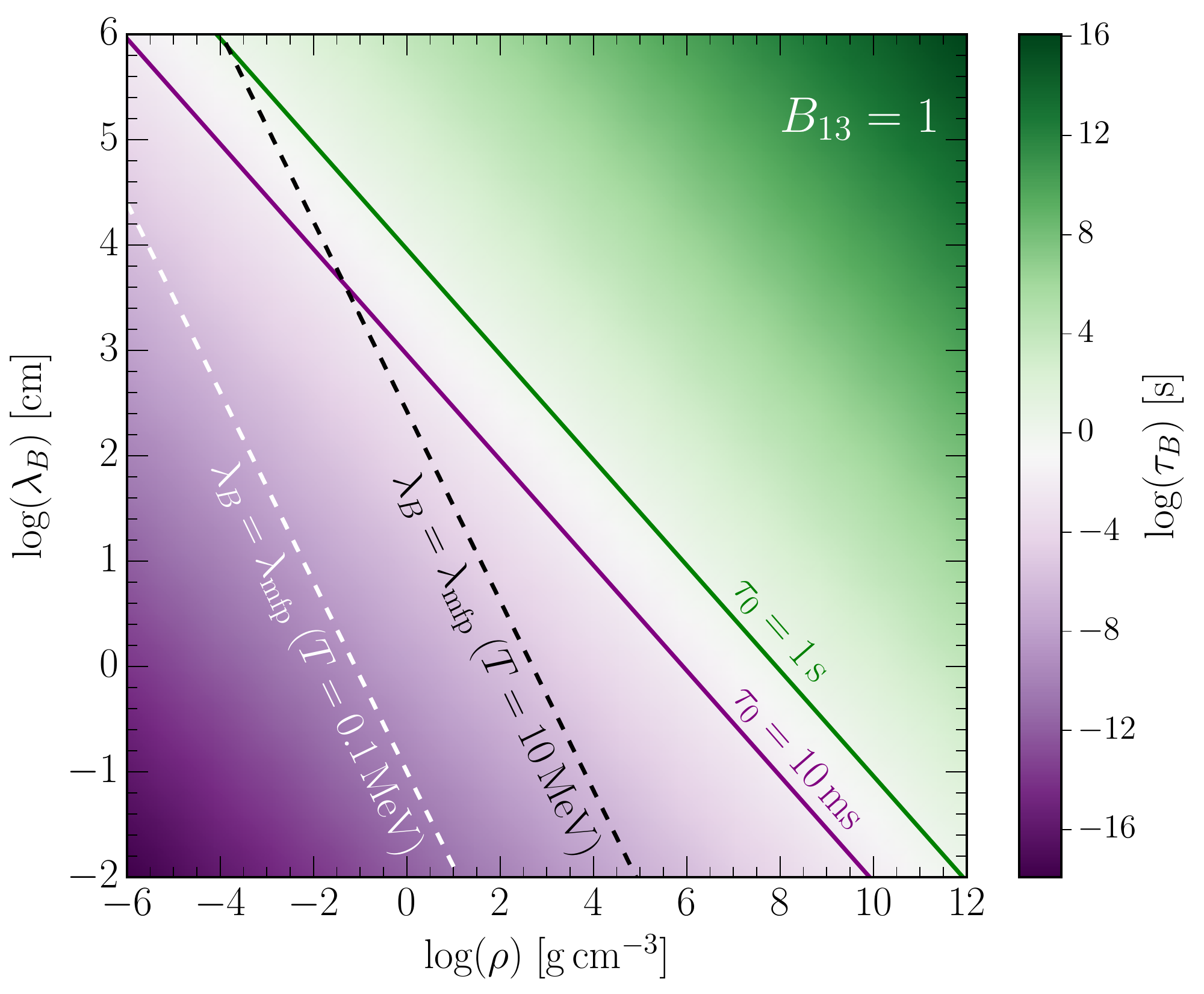}
\caption{Dependence of the Hall timescale $\tau_B$ on the rest-mass
  density and the typical scale-height of the magnetic field
  $\lambda_B$.  The solid lines correspond to typical timescales of
  $\tau_0=10\,{\rm ms}$ and $\tau_0=1\,{\rm s}$, respectively. The
  region with $\lambda_B\geq \lambda_{\rm mfp}$ and $\tau_B\leq\tau_0$
  is where the Hall effect becomes important. The dashed lines show
  where $\lambda_B=\lambda_{\rm mfp}(\rho,T)$ for $T=0.1\,{\rm MeV}$
  and $T=10\,{\rm MeV}$; below these lines the MHD description breaks
  down.}
\label{fig:Hall_time_2D}
\end{center}
\end{figure}

We can make similar considerations also for the timescale $\tau_B$
associated with the Hall effect, in which case we can express
Eq.~\eqref{eq:B_time} as 
\begin{eqnarray}
  \label{eq:tau_B1}
  \tau_B \simeq 1.2\times 10^{2}\times B_{13}^{-1}
  \left(\frac{\rho}{1~{\rm g\, cm^{-3}}}\right)
  \left(\frac{\lambda_B}{1~{\rm km}}\right)^2~{\rm s}\,,\quad
\end{eqnarray}
where $B_{13}:=B/(10^{13}~{\rm G})$. Clearly, the Hall timescale $\tau_B$
is much shorter than the diffusion times and it can be in the relevant
range of milliseconds for low densities and large magnetic field. The
transition from the dissipation regime to the Hall diffusion regime,
namely, when the condition $\tau_B< \tau_d$ is fulfilled, occurs at
densities $\rho\lesssim 10^{11} B_{13}\ {\rm g\, cm}^{-3}$ and is
independent of the lengthscale, since both timescales are proportional to
$\lambda_B^2$.

The dependence of $\tau_B$ on the rest-mass density and $\lambda_B$ is
shown in Fig.~\ref{fig:Hall_time_2D} for a fixed value of the magnetic
field $B_{13}=1$. At low densities, \eg
$\rho\lesssim 10^{10}\,{\rm g\, cm}^{-3}$, we have $\tau_B< \tau_d$
for $B_{13}\geq 0.1$. In this region, therefore, the magnetic field
can also be subject to the Hall effect, which will change its
distribution on a timescale $\tau_B$. As seen in
Fig.~\ref{fig:Hall_time_2D} and from Eq.~\eqref{eq:tau_B1}, $\tau_B$
reaches the value $\tau_0=10\,{\rm ms}$ for $\lambda_B=1\,{\rm km}$ at
very low densities $\rho\leq 10^{-4}\, {\rm g\, cm}^{-3}$, while for
$\lambda_B=1\,{\rm m}$ the value $\tau_B = 10\,{\rm ms}$ is reached
already at the densities $\rho\leq 10^{2}\, {\rm g\, cm}^{-3}$, in
agreement with the scaling $\tau_B\propto\rho \lambda^2_B$.

\subsection{Validity of the MHD approach} 
\label{sec:MHD_val}

After having assessed the ranges of validity of ideal MHD, we can now
turn to the next natural question: what limits the applicability of
the MHD approach in the present context?  Clearly, at very low
rest-mass densities the mean free path of particles becomes large and
the validity of the MHD description of matter itself can break
down. We recall that hydrodynamic description of matter breaks down
whenever the electron mean free path
$\lambda_{\rm mfp} \geq \lambda_B$.

 The mean free path of electrons is
defined as $\lambda_{\rm mfp}=\tau v$, where $\tau$ is the mean
collision time. For nondegenerate electrons, $v^2=3Tc^2/\varepsilon$
so that we obtain [see the first equation in~\eqref{eq:sigma_drude}]
\begin{eqnarray}
  \label{eq:mfp}
  \lambda_{\rm mfp} = \frac{\sigma\varepsilon v}{n_ee^2c^2}
  =\frac{Am_n}{Z e^2c}\frac{\sigma}{\rho}\sqrt{3T\varepsilon}\,.
\end{eqnarray}
In essence, at low rest-mass densities and large temperatures, \ie for
$\rho_6\ll1$ and for temperatures $T\gtrsim 1\,{\rm MeV}$, we can
obtain a simple estimate for Eq.~\eqref{eq:mfp} [the arguments are similar to those leading to Eq.~\eqref{eq:decay_time2a}]
\begin{eqnarray}
  \label{eq:mfp1}
\lambda_{\rm mfp}\simeq 
4.2\, \left(\frac{\rho}{1~{\rm g\, cm^{-3}}}\right)^{-0.9}
\bigg(\frac{T}{1~{\rm MeV}}\bigg)^{1.8}~{\rm cm}\,,
\end{eqnarray}
where we approximated $\varepsilon\simeq 3T$.
 Now we can rewrite the
condition of applicability of MHD description, \ie $\lambda_{\rm mfp}\le
\lambda_B$, as
\begin{eqnarray}
  \label{eq:MHD_range}
\rho \gtrsim  1.4\times 10^{-5} 
\bigg(\frac{T}{1~{\rm MeV}}\bigg)^{2}
\left(\frac{\lambda_B}{1~{\rm km}}\right)^{-1.1}~{\rm g\, cm^{-3}}\,.\quad
\end{eqnarray}
It follows that the higher the temperature or the smaller the typical
lengthscale $\lambda_B$, the higher the rest-mass density below which the
MHD description breaks down. In Figs.~\ref{fig:decay_time_2D} and
\ref{fig:Hall_time_2D} we show with dashed lines where this happens, \ie
where $\lambda_B=\lambda_{\rm mfp}(\rho,T)$, for two values of the
temperature.

\begin{figure}[tb] 
\begin{center}
\includegraphics[width=0.95\columnwidth]{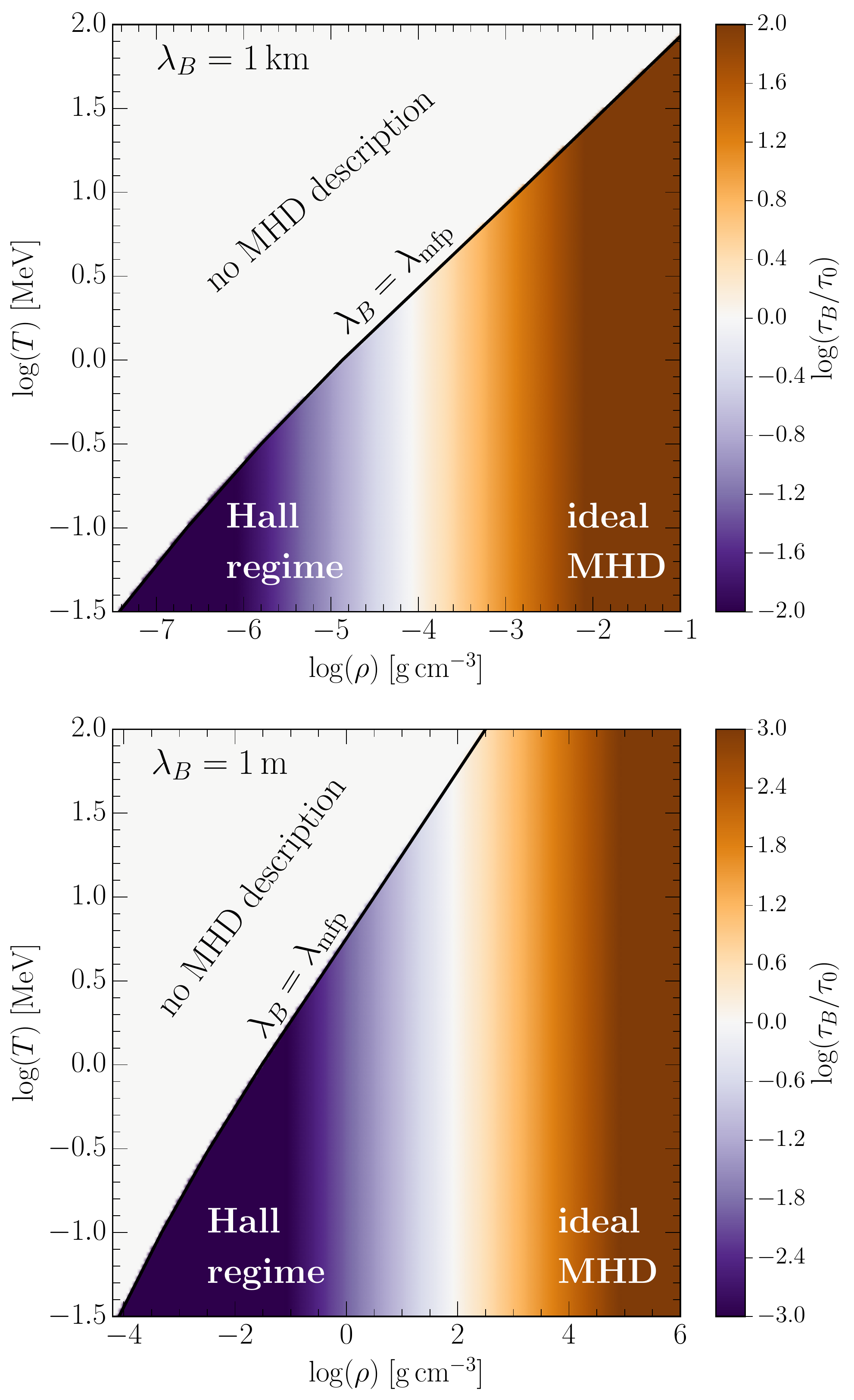}
\caption{Regions of the validity of MHD and ideal MHD on the
  temperature-density plane for two values of the magnetic-field
  scale-height $\lambda_B=1\,{\rm km}$ (\textit{upper panel}), and
  $\lambda_B=1\,{\rm m}$ (\textit{lower panel}). Areas shaded in
  dark-orange are the regions where the ideal-MHD approximation holds;
  areas shaded in dark-violet are the regions where the Hall effect
  becomes important. The value of the magnetic field is fixed at
  $B_{13}=1$, and the typical timescale is taken $\tau_0=10\,{\rm
    ms}$. Above the solid line $\lambda_B=\lambda_{\rm mfp}$ the MHD
  description breaks down.}
\label{fig:ideal_MHD}
\end{center}
\end{figure}

Figure~\ref{fig:ideal_MHD} shows the regions of validity of the MHD
description of matter in the temperature-density plane for
$\lambda_B=1\,{\rm km}$ and $\lambda_B=1\,{\rm m}$, assuming matter as
being composed of hydrogen (the case of matter composed of iron is
discussed below). The temperature range considered is limited from
below by the solidification of matter and covers the range
$-1.5\le \log(T)\le 2$. The low-temperature and high-density region
corresponds to the regime where ideal MHD conditions are
fulfilled. Adjacent to this, a lower density region emerges where the
MHD is applicable, but the Hall effect should be taken into account;
the exact location of this region depends on the strength of the
magnetic field. It moves to lower rest-mass densities for the weaker
magnetic field. Above the separation line
$\lambda_{\rm mfp}=\lambda_B$ (solid black line in
Fig.~\ref{fig:ideal_MHD}), the low-density and high-temperature region
features matter in the non-hydrodynamic regime, \ie in a regime where
the MHD approximation breaks down and a kinetic approach based on the
Boltzmann equation is needed [for a discussion of kinetic regime in
the context of relativistic MHD see \cite{Rezzolla_book:2013}].

We note that due to the constraint on the rest-mass density given
by Eq.~\eqref{eq:MHD_range}, the decay timescale $\tau_d$ is bounded from
below by
\begin{eqnarray}
  \label{eq:decay_time3}
  \tau_d 
   \gtrsim 3.3\times 10^{5}\,\,
  \bigg(\frac{T}{1\,{\rm MeV}}\bigg)
  \left(\frac{\lambda_B}{1\,{\rm m}}\right)^{1.9}~{\rm s}\,.
\end{eqnarray}
Assuming now that the merger process is characterized by extremely low
values of $\lambda_B$ and $T$, \eg $\lambda_B=0.1\,{\rm m}$ and
$T=0.01\,{\rm MeV}$, we find from Eq.~\eqref{eq:MHD_range} that the
MHD description is valid down to a rest-mass density
$\rho_{\rm min} \simeq 3.5\times 10^{-5}~{\rm g\, cm^{-3}}$. In this
case, Eq.~\eqref{eq:decay_time3} gives an estimate
$\tau_d\simeq 40\,{\rm s}$, which is still two orders of magnitude
larger than the typical timescale $\tau_0\sim 10\,{\rm ms}$. Hence, we
conclude, that \textit{the ideal-MHD approximation in binary
  neutron-star simulations is always justified as long as the MHD
  description itself is valid}. Again, we stress that our conclusion
is in contrast to the existing paradigm of modelling the
conductivities in neutron star binary mergers, where low-density
matter features low conductivity and requires dissipative MHD
treatment.  
 
In closing, we ask the inverse question: given that the merger process is
characterized by the timescale $\tau_0=10\,{\rm ms}$, at what
characteristic scales, $\lambda_{\rm lim}$, do resistive effects
become important? To answer this question, we consider $\tau_d\lesssim
10\,{\rm ms}$ in Eq.~\eqref{eq:decay_time3} to obtain 
\begin{eqnarray}
\label{eq:l_lim}
\lambda_{\rm
    lim} \simeq 0.1\,\, \left(\frac{T}{1\,{\rm MeV}}\right)^{-0.5}~{\rm mm}\,.
\end{eqnarray}
 Thus,
in order for resistive effects to be relevant on the typical timescales
of the order of 10 ms and for typical temperatures $T\gtrsim 1\,{\rm
  MeV}$, the characteristic lengthscales of the problem should be of the
order of (or less than) a tenth of a millimeter.

For completeness, we also comment on how our estimates will change if
the matter consists of heavier elements, \eg iron nucleus, which has
$Z=26$. As seen from Eqs.~\eqref{eq:decay_time}, [see
Eqs.~\eqref{eq:sigma_fit}--\eqref{eq:fit_constants2}], the
$Z$-dependence in the timescale $\tau_d$ arises mainly from the
scaling $\tau_d\propto Z^{-1}$. As a result, in the case of matter
composed mainly of iron, $\tau_d$ will be reduced roughly by an order
of magnitude. On the other hand, the Hall timescale $\tau_B$ depends
on the type of nuclei as $\tau_B\propto Z/A\simeq 0.5$ for $Z>1$ [see
Eq.~\eqref{eq:B_time}].  Hence, $\tau_B$ is smaller by a factor of two
in the case of iron when compared to hydrogen and our arguments on the
applicability of ideal MHD remain qualitatively valid also when
$Z=26$.

\section{Conclusions} 
\label{sec:conclusions}

The multimessenger astronomy era started with the observations
of GW170817 in electromagnetic and gravitational waves. Such events,
among many other implications, already place constraints on properties
of compact star integral parameters, such as masses, radii, and
deformability~\cite{Bauswein2017b, Margalit2017, Radice2017,
  Rezzolla2017, Ruiz2017, Shibata2017c, Paschalidis2017, Most2018} and
further insights in various aspects of compact stars are anticipated.
In parallel to these developments, the assessment of the transport and
dissipation in these events has been brought into focus of
microphysics research recently~\cite{Alford2018,Harutyunyan:2016a}.

Motivated by these recent developments, we addressed in this work the
role of dissipative processes in the MHD description of neutron star
binary mergers. Using recently obtained conductivities of warm plasma
in the low-density matter, we have analysed the timescales for the
evolution of the magnetic field under the conditions which are
relevant for binary neutron-star mergers. We found that the
magnetic-field decay time is much larger than the relevant timescales
for the merger process in the entire density-temperature range
characteristic for these processes. In other words, the ideal MHD
approximation is applicable throughout the entire processes of the
merger. This conclusion holds for lengthscales down to a meter, which
is at least an order of magnitude smaller than currently feasible
computational grids. Our finding implies a paradigm shift in resistive
MHD treatment of binary neutron star mergers, as these were based on
the modeling of conductivities which vanish in the low-density
limit~\cite{Dionysopoulou2015}. We have demonstrated that the ideal
MHD description does not break down due to the onset of dissipation,
rather it becomes inapplicable when the hydrodynamic description of
matter becomes invalid.

We have demonstrated that the Hall effect plays an important role in
the low-density and low-temperature regime. Thus, the ideal MHD
description must be supplemented by an approach which takes into
account the anisotropy of the fluid via the Hall effect. As is
well-known, the Hall effect can act as a mechanism of rearrangement of
the magnetic field resulting in resistive
instabilities~\cite{Gourgouliatos2016,Kitchatinov2017}.

While our study suggests that the ideal MHD is valid for the binary
neutron star simulations, a definitive answer can be reached only
through fully numerical studies which include all possible dissipative
effects, namely, finite electrical and thermal conductivities, as well
as shear and bulk viscosities. Luckily, observations from merging
binary neutron stars in the coming years will also provide useful
information to resolve this issue.

Special relativistic dissipative MHD formulations including the full
anisotropy of the transport coefficients in magnetic fields were
developed previously \cite{Huang2010,Huang2011,Hernandez2017}.  A
general-relativistic formulation of the binary neutron star merger
problem which includes the anisotropies of the electrical conductivity
has been given already by \cite{Dionysopoulou2015}. We hope that our
study will stimulate simulations which will include these effects.

\section*{Acknowledgements}  
AH acknowledges support from the HGS-HIRe
graduate program at Goethe University, Frankfurt. AN is supported by an
Alexander von Humboldt Fellowship. AS is supported by the Deutsche
Forschungsgemeinschaft (Grant No. SE 1836/4-1). Support comes also in
part from ``NewCompStar'' and ``PHAROS'', COST Actions MP1304 and
CA16214; LOEWE-Program in HIC for FAIR; European Union's Horizon 2020
Research and Innovation Programme (Grant 671698) (call FETHPC-1-2014,
project ExaHyPE), the ERC Synergy Grant ``BlackHoleCam: Imaging the
Event Horizon of Black Holes'' (Grant No. 610058). The simulations
were performed on the SuperMUC cluster at the LRZ in Garching, on the
LOEWE cluster in CSC in Frankfurt, on the HazelHen cluster at the HLRS
in Stuttgart.

\input{Bdecay_EPJA.bbl}

\end{document}

%% file: Bdecay_EPJA.bbl
\providecommand{\href}[2]{#2}\begingroup\raggedright\endgroup